\documentclass[]{revtex4}

\usepackage{graphicx}% Include figure files
\usepackage{dcolumn}% Align table columns on decimal point
\usepackage{amsmath,amsfonts}% popular packages from the American Mathematical Society
\usepackage{bm}% Bold Math package
\usepackage{color}
\usepackage{subfigure}

\begin{document}

\title[Ribs on Soundboard]{The effect of Mounted Ribs on the Radiation of a Soundboard}
\author{Marcel Kappel}
 \email{makappel@uni-potsdam.de}
 \altaffiliation[]{UP Transfer GmbH at the University of Potsdam}%Lines break automatically or can be forced with \\
\author{Markus Abel}%
 \affiliation{University of Potsdam, Institute of Physics and Astronomy}%Line breaks may be forced with \\ here, too
\author{Reimund Gerhard}%
 \affiliation{University of Potsdam, Institute of Physics and Astronomy}%Line breaks may be forced with \\ here, too

\begin{abstract}
The grand piano is one of the most important instruments in western music. Its functioning and details are investigated and understood to 
a reasonable level, however, differences between manufacturers exist which are hard to explain. To add a new piece of understanding, we 
decided to investigate the effect of ribs mounted on a soundboard. Apart from pianos, this is important to a wider class of instruments 
which radiate from a structured surface.
From scattering theory, it is well-known that a regular array of scatterers yields 
a band structure. By a systematic study of the latter, the effect of the ribs on the radiated spectrum is demonstrated 
for a specially manufactured multichord mimicking topologically a piano soundboard.
To distinguish between radiated sound and sound propagated inside the board we use piezopolymers, an innovative, non-invasive technique.
As a result we find a dramatic change in the spectrum allowed to propagate in the soundboard which is consequently radiated. An 
explanation by a simple model of coupled oscillators is given with a very nice qualitative conincidence.
\end{abstract}


\maketitle

\section{Introduction}
\label{sec:intro}
The physics of pianos was always subject to intense research on musical
instruments. According to common scientific understanding, sound is generated and radiated in a chain of processes:
the hammering mechanism
\cite{Askenfelt-Jansson_series,Hall_series,Chaigne-Askenfelt_series}, the
vibration of coupled strings \cite{Weinreich-77}, the sound transmission of
vibrations through bridge and soundboard
\cite{Conklin_series,Giordano-Korty-96,Giordano-Gould-Tobochnik-98}.  Sound
radiation from a resonance board is described in
\cite{Moore-Zietlow-06,Conklin_series,Giordano-98,Suzuki-86}, some models
of piano soundboards have been proposed
\cite{Giordano-97,Bensa-Bilbao-Kronland-Martinet-Smith-03}. Large
boards are typically manufactured using so-called ribs as a support; they have
a double function in the instrument. On one hand, the ribs enhance the
mechanical stability of the instrument, on the other hand they pre-stress the
soundboard and thus act as an energy reservoir ready to release in case.  The
effect of this mechanical construction, on the rear side of the resonance
board has been mainly neglected in the literature.

Ribs are typically arranged in a regular array, reminiscent to scatterers in a crystal. From solid state theory \cite{Ziman-72,Kittel-53,Ashcroft-Mermin-76}, we know that such an arrangement produces particular eigenmodes and -frequencies. The latter are recognized in a band structure with ``allowed''
frequencies in a certain band given by the scattering properties of scatterers (atoms or molecules in solids). We map this problem to 
the problem of a sound wave propagating in the sound board. Individual scatterers are represented by the ribs. The soundboard can be considered as 
a waveguide with varying cross section due to ribs and other mounted parts (in particular the bridge). Since we have a highly finite system, in principle
we must take into account the boundary conditions (bc), 
which are basically unknown, since the boundary is neither fixed nor open (mixed Dirichlet-von Neumann bc).

The investigation of the effect is difficult when the 
particularly shaped sound board of a grand piano is considered: the eigenmodes
depend heavily on the specific shape. However, 
a surface structured with regular ribs can be mapped to the soundboard, 
e.g. by conformal mapping. Under mild restrictions, the structure of the 
frequency spectrum will be conserved.
Consequently, we decided to build a dedicated setup 
by a rectangular multichord which is easily understood,
cheap and well controllable. The multichord is fully equipped with boundary bridges (for the defined string length), the excitement bridge, tuning pegs and piano strings. \\
The multichord was used to mount the ribs and measure the change of the radiated sound and the sound propagated in the board itself.
The former is measured by a conventional microphone, the latter by state-of the art piezopolymer films, described in detail in \cite{Kappel-Abel-Gerhard-11} 
and to some extent below. The main advantage of such films is their extremely thin design, being just a few microns in height, their extension is
macroscopic (sub-millimeter to centimeter size). Consequently such a film is placed between the sound bridge and the sound board. The jump in impedance
due to this microscopic obstacle can be safely neglected.

Our results confirm fully the intuition: mounting an increasing number of ribs yields a suppression of large regions in frequency spectrum.
We present detailed measurements on the number of ribs mounted, the distance and the shape of the spectrum. A qualitative explanation is
found by one-dimensional theory of coupled oscillators. Quantitatively, a match is not straightforward, since neither the material parameters
can be assumed to be given, nor do we know the boundary conditions to a sufficient detail. 

This publication is structured as follows. After this introduction we give details on the experimental setup and measurement techniques.
In Section \ref{sec:theory}, we outline the theory of scattering and resulting eigenmodes in a finite-size array with different distances between 
the scatterers. This is followed in Sec. \ref{sec:results} by the presentation of the main results and their explanation. Finally, we
conclude with Sec.~\ref{sec:conclusions}.

\section{Experimental Setup}

\begin{figure}[ht]
\subfigure{(\textbf{a})}{\includegraphics[width=0.47\textwidth]{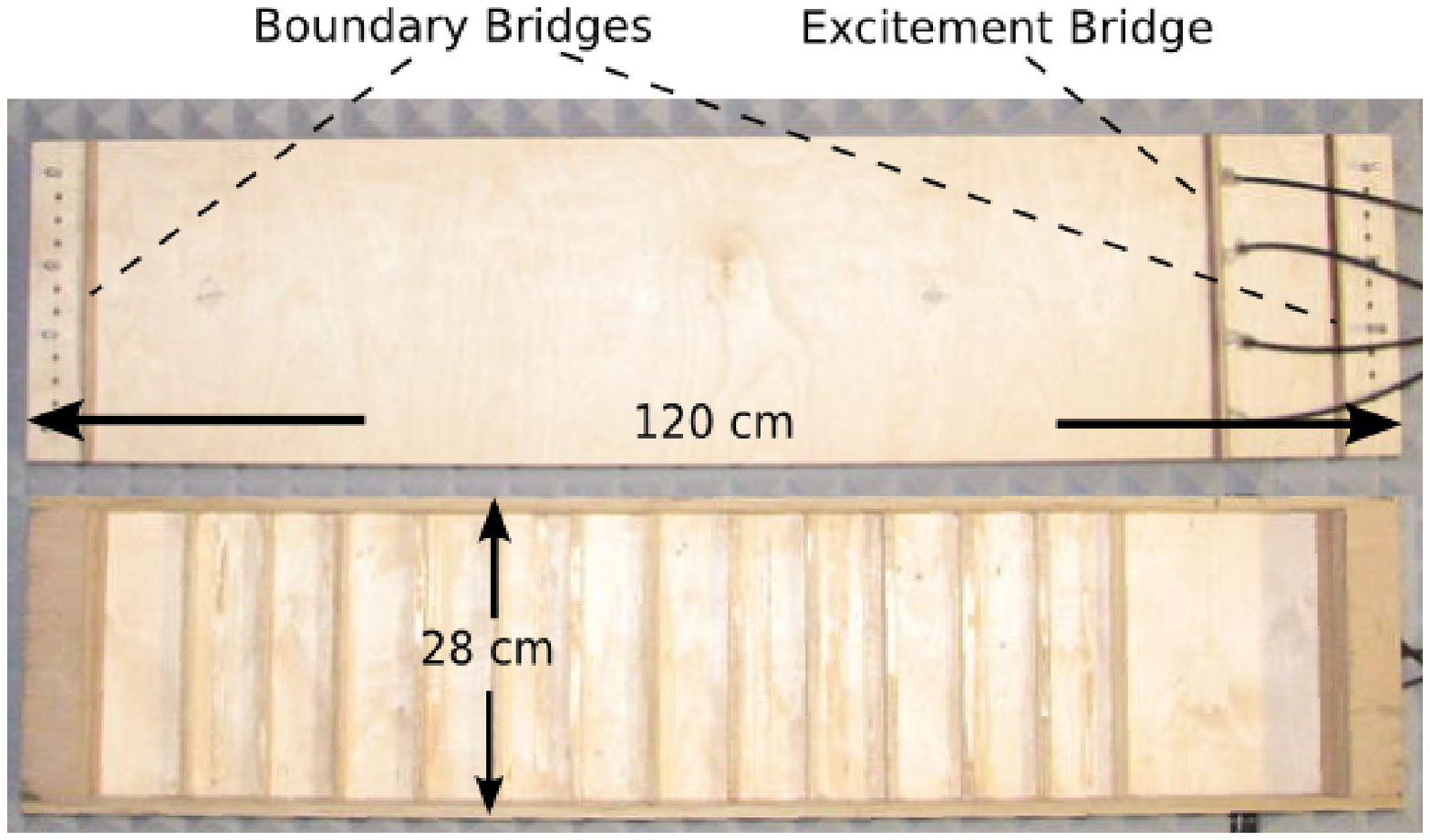}} % pictures of the multichord with dimensions
\subfigure{(\textbf{b})}{\includegraphics[width=0.47\textwidth]{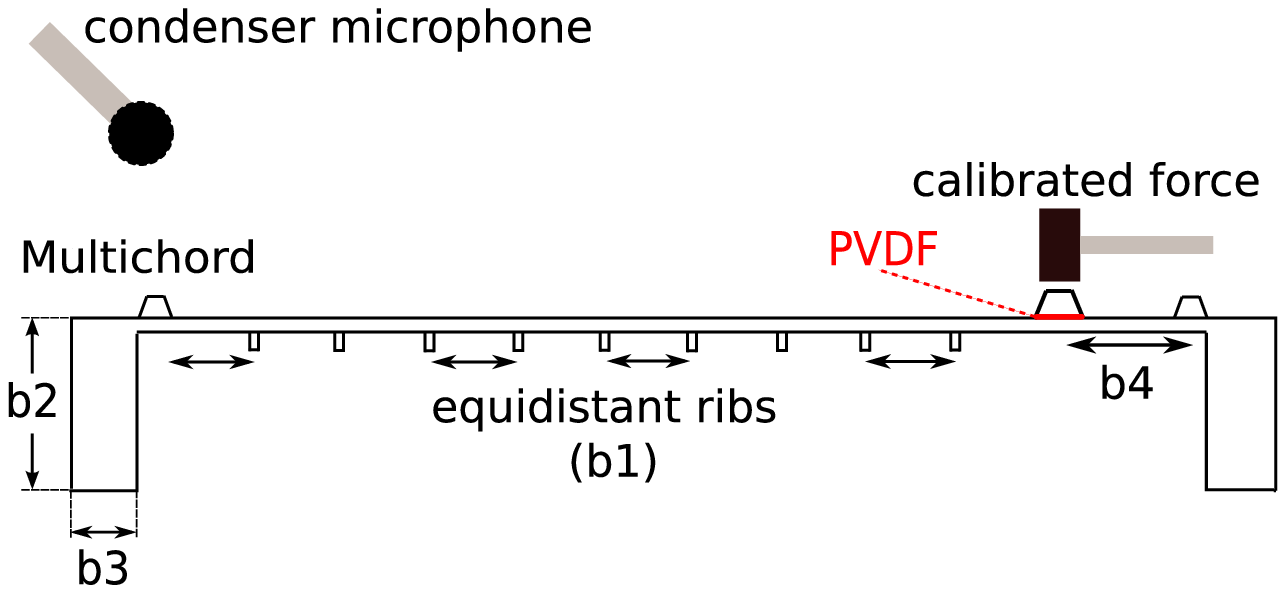}} % multichord configuration sketch sideview
\caption[Multichord sketch]{\textbf{a:} Top and bottom view of the multichord 
without mounted strings and with ribs on the rear side of the soundboard. 
The measurements were performed without strings. \\
\textbf{b:} Sketch of the multichord setup with ribs mounted on the rear side of the sound board (\textbf{b1} denotes the distance between ribs). The hammering device is sketched on the right hand side and excites the resonance board via the sound bridge with a defined impulse. The acoustic waves were recorded by a B\&K condensor microphone above the arrangement and by a piezopolymer mounted between the bridge and the board. Further dimensions are: height \textbf{b2} = 7.8 cm, boundary thickness \textbf{b3} = 5 cm and disctance the excitement bridge -- boundary bridge, \textbf{b4} = 11 cm.}
\label{fig:multichord}
\end{figure}

The main objective of this research is the investigation of the spectral characteristics of the sound propagated in and radiated from a soundboard with
mounted ribs. We constructed a rectangular shaped multichord with bridge and other string mountings, under the soundboard we glued a number of 
equidistant ribs and successively  measured their effect, 
cf. Fig.~\ref{fig:multichord}. The ribs were placed such that their distance to the 
(above mounted) bridge equalled the inter-rib distance. The dimensions of the soundboard are 1.2 x 0.28 m, the wood used for the soundboard was conventional 
blockboard maple glued in alternating layers. 
This is, of course, not so in a grand piano where massive wood with very special properties is used. But this way we avoid
anisotropy along length and width, except for the mountings we add by ourselves. For musical purposes we mount piano strings as used by Bechstein \cite{Bechstein-05}. After some experiments where strings were excited by controlled hammering, it turned out that an excitation mechanism with 
a mechanical hammer hitting directly the bridge were much more convenient, since we excite all relevant frequencies at once.
To have a sound base, the hammering amplitude was measured in a series of consecutive experiments. As a result, we found that the amplitude was constant
with slight deviations. For our experiments, the focus was on frequency analysis and we did not further discuss about the 
amplitude of the hammering, anticipating to have a LTI system (linear time invariant). 

The ribs were placed with equal distances of 7, 9.8, and 14 cm, starting from the bridge. This way, at one side (\textbf{b4}) one non-equidistant length exist, cf. 
Fig.~\ref{fig:multichord}. The experimental procedure was the following. We start with an ``empty'' soundboard which is the same for all three 
series. Then step by step we increase the number of ribs by one until the board is filled to the maximum number of ribs possible (13, 9, and 6 for 7,9.8, and 14 cm respectively).

For simultaneous measurement, we used a calibrated piezopolymer in combination with a condenser microphone (Br\"{u}el \& Kj\ae r Free-field 1/2`` condenser microphone (Type 4191) with Br\"{u}el \& Kj\ae r amplifier Nexus Type 2691), cf. Fig.~\ref{fig:multichord}.
The piezopolymer was mounted between the sound bridge and the resonance board with a defined pressure of 55 kPa. The B\&K condenser microphone, placed above the resonance board at an angle of $32^\circ$ with respect to the middle of the resonance board to imitate the position of the audience in a theatre hall, measured the overall airborne sound, cf. Fig.~\ref{fig:multichord}.

\begin{figure}[ht]
\subfigure{(\textbf{a})}{\includegraphics[width=0.47\textwidth]{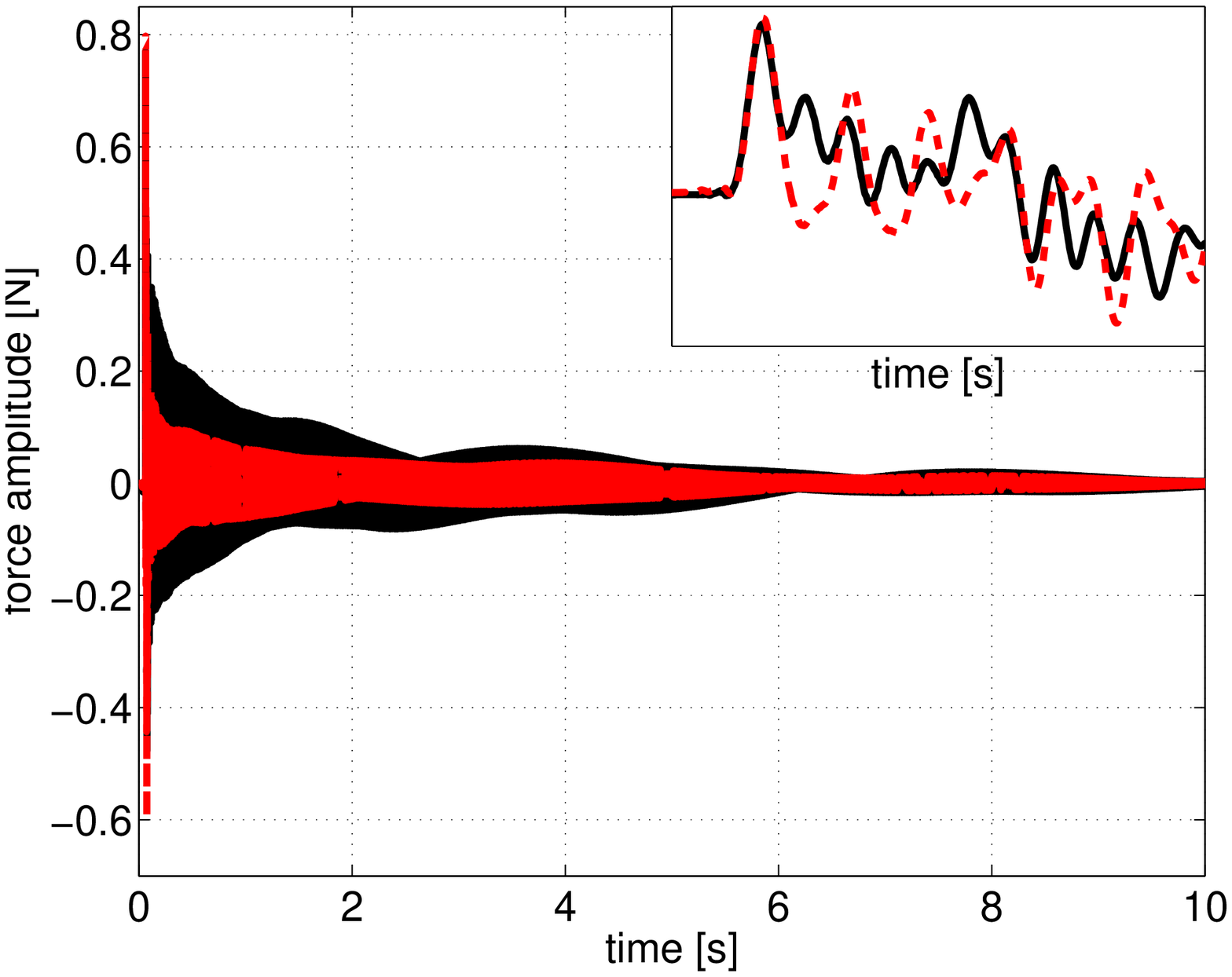}} % decay of the envelope for an example measurement (plot_decaytime_tiled_RB  first part)
\subfigure{(\textbf{b})}{\includegraphics[width=0.47\textwidth]{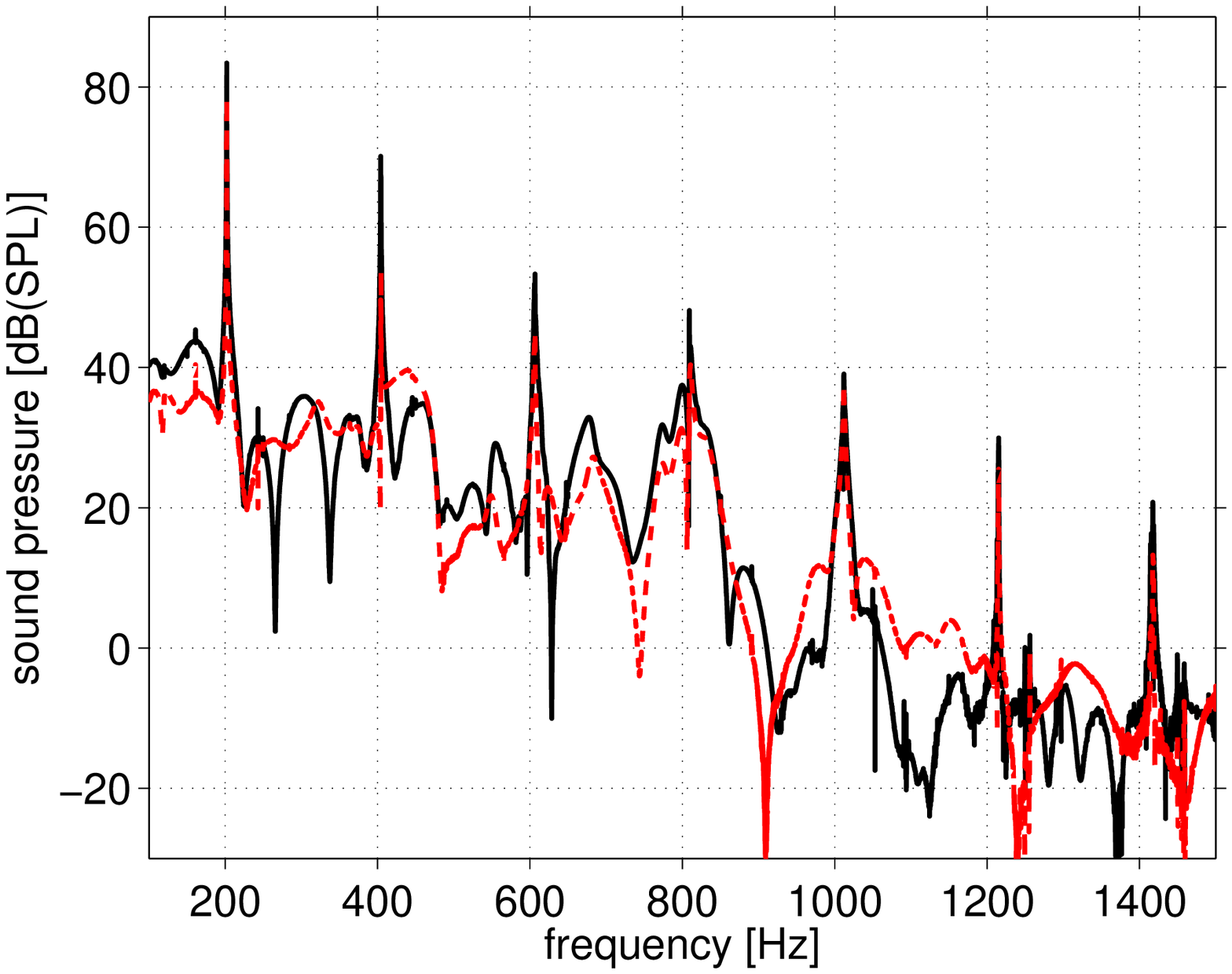}} % frequency of these tones (plot_decaytime_tiled_RB  second part)
\caption[Piezopolymer measuring example]{Piezopolymer measuring example. \textbf{a}: Sample of different excitation positions on a mounted string: excitation at one-third of the string (black), and excitation at the half of the string (red). The inset shows the initial force for either way. The curves coincide well demonstrating the reproducibility of the measurements. \\
\textbf{b}: Frequency spectra in the range of 100 Hz to 1500 Hz of both example time signals. 
The differences in peak heights are clearly due to the different string excitation.}
\label{fig:decay_behav}
\end{figure}

Since the piezoelectric polymer material was manufactured especially for the soundboard measurements, a short description of its main characteristics is given here. A detailed account of the technical properties is given in \cite{Kappel-Abel-Gerhard-11}. We carefully checked the properties of the material used and calibrated the measurements in order to ensure comparability. The film material used in the soundboard is mono-axially stretched pre--poled piezoelectric polyvinylidene fluoride\cite{Kawai-69} manufactured by Piezotech S.A.S. with a piezoelectric coefficient of $d_{33_{\mathrm{PVDF}}} = 16.3$pC/N. All samples have a thickness of $30 \mu$m and a density of $\varrho = 1,77\mathrm{g/cm^3}$.
The frequency dependence of the piezopolymer was checked with a sweep method in the range of 100Hz to 17000Hz. We find a non-linear frequency response, which is a known effect for sensors with capacitance. Theory on the frequency transmission is described in \cite{Rabiner-Gold-75}. All measured piezopolymer signals were corrected by the response curve. The piezopolymer signal-to-noise ratio at the first harmonic of the multichord is $L_p = 78$dB.

We carried out a series of experiments: To ensure reproducibility of the measurement signal,  
a mounted string was excited at two different positions. The comparison of the signals,  measured by the polymer film, is shown in Fig.~\ref{fig:decay_behav};  (\textbf{a})  time signal, and (\textbf{b}) frequency spectrum . The initial displacement is the same for both positions and show that the deviations of the signal are small (inset in \textbf{a}).

For 13 ribs mounted with distance 7 cm, but  without strings we repeated the experiment 23 times and found no significant spectral differences, in particular, the position of the peaks was identical.

\begin{figure}[ht]
\includegraphics[width=0.47\textwidth]{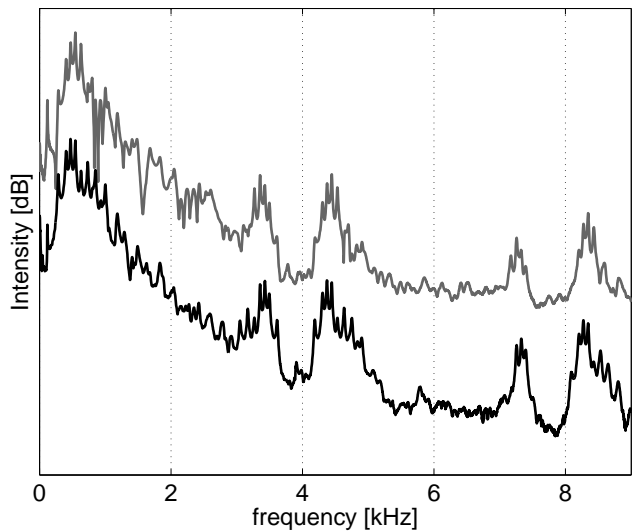} % the difference from different exciting points
\caption[Different excitement points]{
Spectral differences for different excitation points. The gray curve represents the bridge excitation (used throughout this publication), the black line is the outcome of a decentral excitation in the middle of a void space between two ribs. The curves are separated by a factor of 100 for a better comparison. Whereas attack at the bridge shows some structure off the bands of interest here, the attack directly on the soundboard is a bit smoother.  Since it is musically relevant, we use the bridge excitation and note that the band structure of interest is basically identical.}
\label{fig:different_excitement_point}
\end{figure}

The point of attack of the hammer was changed from central on the bridge 
to decentral in the middle of a void space between two ribs. There are 
differences, which are displayed in Fig.~\ref{fig:different_excitement_point}.
In principle one could  expect changes, as is known from the attack of 
piano strings, we cannot find deviations larger than
3\% in peak position.

\section{Qualitative Modeling}
\label{sec:theory}
In this section, we want to describe how a model can be developed which 
considers the propagation of waves in the soundboard as a scattering 
problem, where each rib represents a scatterer. Whereas in general, 
the vibrations of a membrane are described by a fourth order wave equation,
the body vibrations follow a usual second order wave equation which we refer to 
below. 
The eigenmodes are dispersive traveling waves \cite{Guyader-06,Hagedorn-DasGupta-07,Landau-Lifshitz}. We now illustrate our ideas
on how the soundboard vibration and subsequent radiation is happening.

Let us consider the asymptotic situation. The wooden plate shall be infitely extended one
direction, and might be bounded laterally. In essence this amounts to 
an idealization towards a quasi-1D system. At zero an infinite
series of ribs with distance $a$ starts. Then, a sound wave incident from $-\infty$ is scattered forward and backward by the first scatterer according to the scattering function, which depends on the wavelength and the relative phase of the incident wave.
The second scatterer does the same, and so on. This procedure is described by the transfer matrix theory \cite{Crisanti-Paladin-Vulpiani-93}.
On the other hand, one can consider the system as a whole and describe the system in one big, infinite matrix, whose eigenvalues and -vectors can be computed. Both procedures can be mapped into each other. Numerically,
the transfer matrix approach has the advantage to need very small memory for simulation. 

Think about a further degeneration: take $N+1$ ribs being infinitely heavy and infinitely thin, the board being finite size of $N\cdot a$.
Then waves cannot travel, and between the ribs only standing waves with a wavelength suiting the resonance condition $\lambda_n=2a/n$, $n\in \mathbb{N}$ are solutions of the system. The according frequencies for an elastic membrane, $\nu_{n,0} = c(\nu)/\lambda_n$, are 
$N$-fold degenerate, $c$ being the speed of sound. 

If now the mass of the ribs is very large but finite, this amounts to a coupling of the former independent waves with a coupling constant, say, $\epsilon$. Consequently the degeneration 
is lifted and the frequency spectrum consists of $N$ separate eigenfrequencies within a band of width $2\epsilon$, which starts at the frequencies $\nu_n$.%, cf. Fig.~\ref{fig:overall_demo}. 
In essence, this amounts to map the extended system with elasticity to a set of coupled oscillators with frequencies $\nu_n$; we neglect the 
 finite speed of sound and assume nearest-neighbor coupling only:
\begin{equation}
 \ddot{u}_j = -\nu_{n,0}^2 + \epsilon(u_{j+1} - 2 u_j + u_{j-1})
\label{eq:band_per}
\end{equation}
with $j=1,\dots,N$. For periodic boundary conditions, and the ansatz $u_j\sim e^{ik_j j/N}$, $k_j=2\pi/\lambda_j$ one obtains the band of allowed frequencies as
\begin{equation}
 \nu_j^2 = \nu_{n,0}^2 + 2\epsilon\sin^2(kj/(2N))
\label{eq:band_frequencies}
\end{equation}
So, for each resonant frequency $\nu_{n,0}$, now a full band of frequencies exists with the dispersion given by Eq.~\ref{eq:band_frequencies}. The modification for fixed boundary conditions is straightforward and yields frequencies shifted within the band; qualitatively, the result remains the same. 
In general, one has to solve the eigenvalue problem
\begin{equation}
 A \vec{u} = \lambda \vec{u}
\end{equation}
with $\lambda=-\nu^2$ $a_{i,i}= -\nu_{n,0}^2 -2\epsilon $, $a_{i,i\pm 1} = \epsilon$, and $\vec{u}=(u_1,\dots,u_N)$ for $i=2,\dots,N-1$, the respective boundary conditions enter in the first and last line entries. A variation in the distances, or in the coupling is easily inserted in the matrix by allowing $\nu_{n,0}$ or $\epsilon$ to vary for each entry. This yields shifts in the position of the eigenfrequencies, and eventually, if the system is fully disordered 
to a destruction of the pass bands and localization \cite{Condat-Kirkpatrick-86,Crisanti-Paladin-Vulpiani-93,Bayer-Niederdraenk-93}.

The above holds of course for positive and negative frequencies, since the situation is fully symmetric, which is reflected by Eq.~\ref{eq:band_frequencies}. Because the problem is linear, this holds for any coupling, but with increasing coupling, the sharpness of the peaks decreases, i.e., the overlap of the waves increases. To adapt this qualitative picture in more detail to the experiment, the ribs must be taken into account as a jump in cross section and the finite width must be considered. Of course, the corrections at the ends of the soundboard must 
be estimated more accurately and the lateral extension must be correctly inserted in the equations in a further step.
This work is subject to future research, here we focus on the experimental results and refer to \cite{Ziman-72} for the general theory.

What do we expect for our experiment, a soundboard with limitations of finite size, imperfect material, finite-size ribs, finite-mass ribs? 
One can expect a strong decrease of the frequencies {\em not} allowed to 
propagate, i.e., with wrong wavelengths. This is so, because each scatterer 
reflects a part of the incident wave in such a way that it 
interferes destructively with itself. As a result, one should observe frequencies lowered in large regions of the spectrum and bands of 
allowed frequencies (``pass bands'') at regions located approximately
in the intervals $[\pm\nu_n;\,\,\pm\nu_n\pm 2\epsilon]$.%, cf. Fig.~\ref{fig:overall_demo}.

However, the soundboard itself possesses an intricate structure of
eigenmodes. The most important one is the first longitudinal mode
determined by the length of the soundboard, $L_{SB}$. The next resonance in
longitudinal direction is given by the distance of the excitement bridge 
to the boundary of the board, $L_{BR}$.
So, any measurement can only show a banded structure 
{\em on top} of the eigenmodes $\nu_{SB}$ of the unperturbed soundboard. 
So, we expect that we find a peak at $\nu_{0,long} \simeq c \cdot 2L_{SB}$ 
plus attached band from the ribs, a further peak at $\nu_{1,lon} \simeq c \cdot 2L_{BR}$, etc. 
In the next section we will see that these predictions are perfectly confirmed. Again, the situation is fully symmetric with respect to left and right traveling waves, such that we should find a mirrored band at $\nu_{SB,n} \pm \nu_{j}$, $j=1,...,N$.

At last in this section, we want to point out an alternative view.
Above, we started with $\epsilon = 0$ (infinite mass of the ribs) and then let $\epsilon$ small. On the other hand, in the experiment the situation is opposite: we start with an unperturbed board, i.e. without any modes and $\epsilon=\infty$. Then, one should find the unperturbed modes -as we do- 
and see rather a split in the frequencies of these modes in terms of the
resonances on top of the resonance, i.e. the x:(L-x) resonance. This is perfectly analogous to the picture above, but a bit closer to the experimental situation. In both cases one ends up with a discrete set of
natural eigenmodes on top of which the modes due to the periodically arranged ribs are located.

% Then, it is clear that the shift of the innerband peaks due to the changed distance is observed by a change of the position of the peaks by a factor x/y, where x and y are the distances. One has to measure $\nu_x-\nu_1$ (e.g.), and $\nu_y-\nu_1$ and then compare the ratio of these differences. We give only the example of ...

\section{Results}
\label{sec:results}
To obtain results on the effect of ribs, it is favorable to investigate changes in one parameter only. The measurement campain allows to compare
for i) sound propagated in the soundboard and radiated sound (piezopolymer vs. condenser microphone), ii)
sound generated for an increasing number of ribs with the same distance, iii) sound generated for an equal number of ribs with different distances.

\begin{figure*}[ht] % name of the matlab program is:  plot_rib_bands_shaded.
\includegraphics[width=0.7\textwidth]{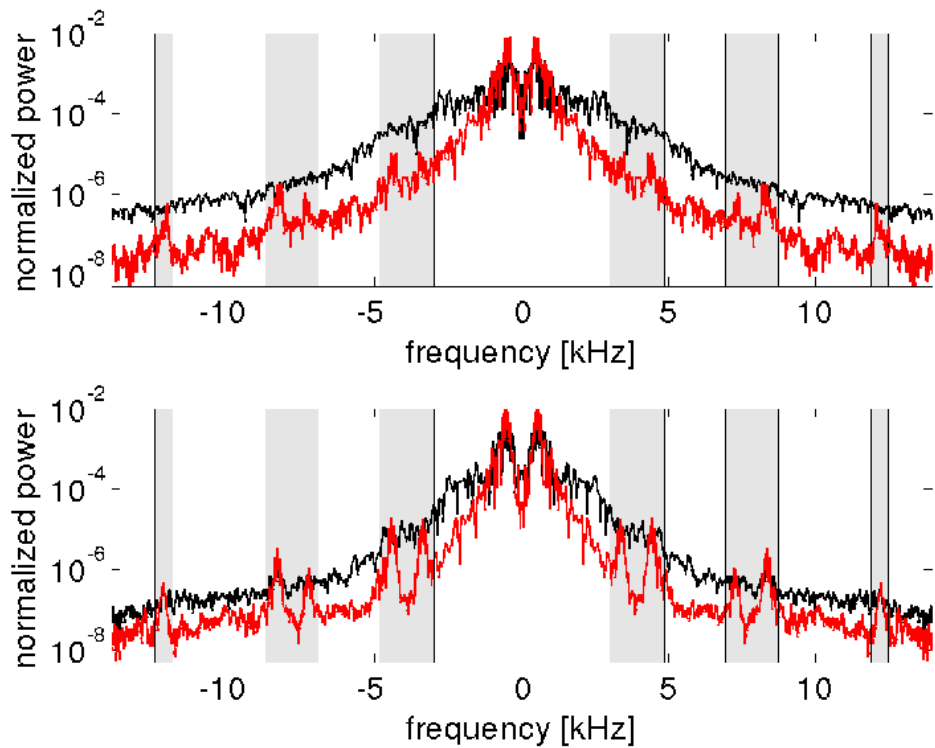} % pass bands equal dist mic  (first part of plot_rib_bands_shaded)
%\subfigure{(\textbf{b})}{\includegraphics[width=0.8\textwidth, height = 0.4\textwidth]{4b.eps}} % pass bands equal dist pvdf (second part of plot_rib_bands_shaded)
\caption[Comparison zero rib spectrum with 13 ribs spectrum]{Spectral comparison: Board without ribs (upper black curve) and with 13 ribs (lower red curve) on the rear side of the resonance board. \textbf{top}: Condensor microphone, \textbf{bottom}:  polymer film  measurements. The shaded regions indicate the positions of the developing passbands. One observes a clear 
banded structure for both signals, but the piezopolymer has a much more pronounced structure with deeper minima. The reason is most likely that radiation acts as a kind of smoothing, damping inside the material reduces the bandwidth of the signal (min  to max ratio). All curves are normalized by the total spectral power.}
\label{fig:overall_demo}
\end{figure*}

Before entering these details, we want to show the comparison of 
a soundboard with many ribs mounted vs. the same one without ribs.
This is shown in Fig.~\ref{fig:overall_demo}a for the microphone,
and in Fig.~\ref{fig:overall_demo}b for the piezopolymer measurement.
The setup was for 13 ribs of 7 cm distance spectra were normalized
to have equal total power. One clearly recognizes the banded structure 
for the soundboard with ribs (red line) in comparison for the empty one (black line). Maybe the most important point is not the existence of 
passing modes, but rather the suppression of the others, which are 
{\em not} allowed to propagate and consequently have a dramatic effect 
on the radiated properties. To our knowledge this is an unprecedented result 
and finding it so clearly can be attributed to our focusing on the ideas of 
wave propagation in a lattice and the idealized setup.
The comparison of piezopolymer and microphone spectra shows that the radiated sound is not as affected as the sound propagated in the board.
However, the band structure is clearly visible and quantifyable.
In the following we analyze the piezopolymer measurements because they 
show more pronounced structures.

%More detail: Fig.~\ref{fig:MicPolyEx}

\begin{figure}[ht] % name of the matlab program is:   plot_multichord_rib_bands_compare_mic_pvdf.m
{\includegraphics[width=0.47\textwidth]{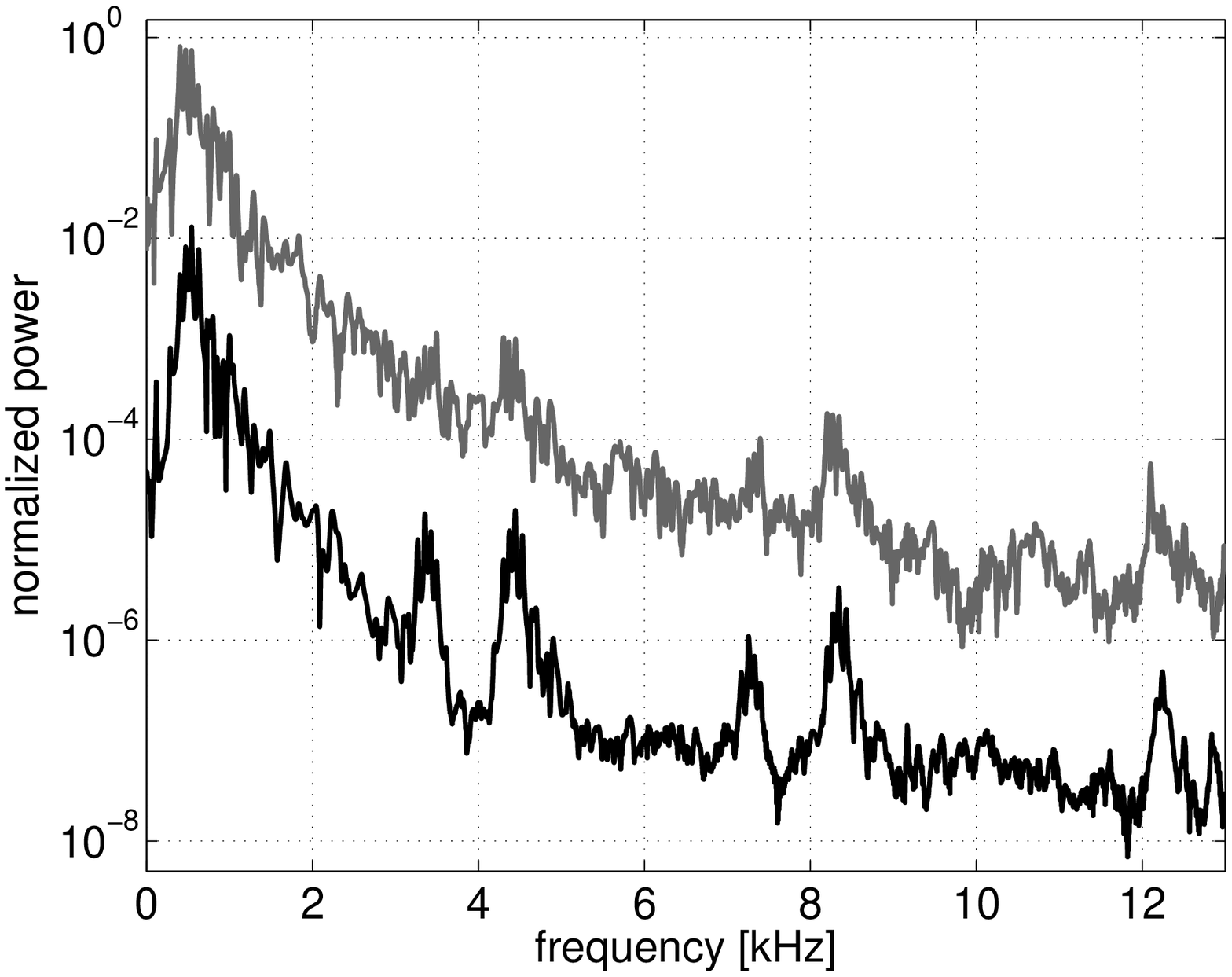}} % comparison mic and PVDF ... whole spec ... 13ribs
\caption[Comparison microphone and piezopolymer spectrum]{Direct comparison of the piezopolymer (black) and microphone (grey) measurements, with 13 mounted ribs. Waves propagating in negative direction in the soundboard are stronger damped, which is reflected by the less expressed ``left hand`` bands in the microphone signals. The curves are normalized and seperated by a factor of 100 for a better comparison.}
\label{fig:MicPolyEx}
\end{figure}

\begin{figure*} % name of the matlab program is:  plot_multichord_rib_bands_shifted_mic_pvdf.m
\includegraphics[width=0.8\textwidth]{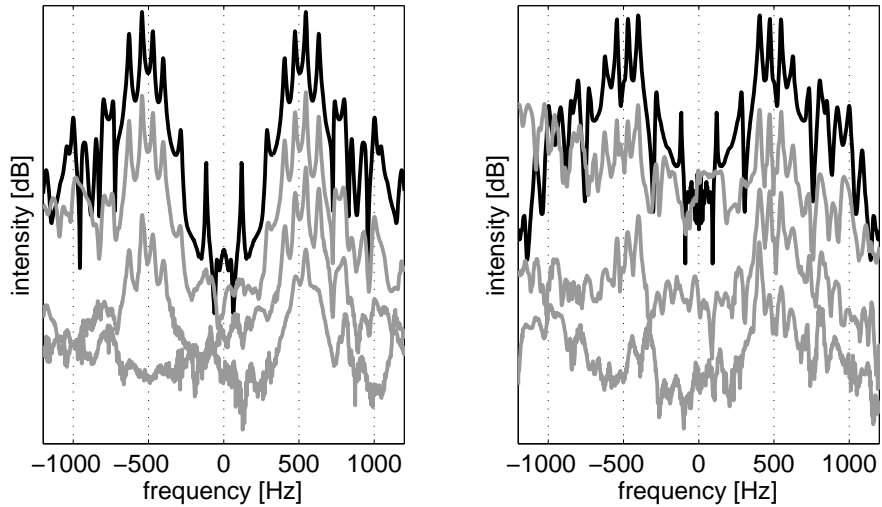} % Shifted frequency range ... mic
%\subfigure{(\textbf{b})}{\includegraphics[width=0.47\textwidth]{6b.eps}} % Shifted frequency range ... pvdf
\caption[Shifted frequency bands]{Comparison of the band structure for several bands, \textbf{right}: microphone , \textbf{left}: piezopolymer.  The first band observed lies in the frequency range of $-1200$ Hz to $1200$ Hz (\textit{black line}). The higher bands are located from $4150$ Hz to $4865$ Hz, $8065$ Hz to $8765$ Hz, and $11950$ Hz to $12665$ Hz, depicted in gray.  The values for the shifts are independent from the number and distance of ribs: 2nd band with $3900$ Hz shift, 3rd band with $7800$ Hz shift, and 4th band with $11700$ Hz shift. The position of the peaks coincides perfectly. The gray curves are shifted towards the black curve by a factor of 100 for a better arrangement.}
\label{fig:shifted_bands}
\end{figure*}

Let us analyze the positions of the band edges, i.e. the underlying mode structure of the board. Without ribs, one should find eigenmodes at 
approximately $\nu_{1,n} = \frac{c}{n\cdot 2 L_{SB}}$ 
for the vibrations along the direction of the strings and 
$\nu_{2,n} = \frac{c}{n\cdot 2 L_{BR}}$ for the mode which resonates
between bridge and boundary. The exact determination of the frequencies of the 
modes is hard to determine, since 
the boundary conditions are not well defined (not fixed, not open), 
and some corrections must be calculated.
We do not know exactly the material parameters, and thus we 
leave the exact determination to the measurement.
The velocity measured in the unperturbed board is $c=773$ m/s, $L_{SB}=1.2$ m, and 
$L_{BR}=0.11$ m; this results in the 
frequencies $\nu_{1,1} =322$ Hz and $\nu_{2,1}=3513$ Hz. We find the first
band in the frequency range of 250 Hz to 965 Hz which is a quite good coincidence,
given the possible errors. From our considerations, we expect another peak for the resonance
of the boundary bridge ($L=0.04$ m); our measurements do not show a clear structure at the expected positions of about $9662$ Hz. We thus should see 12 peaks within a band. Given the relative crude modeling, our findings are 
quite good with 11 peaks found, cf.  Fig.~\ref{fig:shifted_bands}. 
At this point one should note that
especially the frequencies close to the band edges are very sensitive to 
perturbations and can fall out of the band easily; 
then they are heavily damped and no longer seen.

The second band is found from 4150Hz to 4865Hz, the third from 8065Hz to 8765Hz. 
These are {\em exactly} the positions of the first band (negative and positive frequencies), but shifted about $\nu_2= 3900Hz$, we demonstrate this
in Fig.~\ref{fig:shifted_bands} for the distance 7 cm with 13 ribs. 
There, we shifted the second, third, and fourth band for 
positive frequencies by $\nu_1 + m\cdot \nu_2$, $m=2,3,4$. The coincidence is very convincing.
But for negative frequencies $\nu_1 - m\cdot \nu_2$, the corresponding band disappears for $m=4$,
and is already less prominent for $m=3$. The reason is the following:
Higher frequencies are stronger damped, consequently waves propagating in negative direction (represented by ``negative frequencies'') are damped out. They are typically scattered many times
 and thus have run a long way before returning to the measurement point, i.e the piezo device under the excitement bridge.
Concluding this paragraph: one finds bands starting at positions $\pm \nu_1$, and $m\cdot \nu_2\pm \nu_1$. The band structure is exactly repeated for the ``higher'' bands, i.e. the ones found at $m=2,3,4$. The last structure is observed from 11950Hz to 12665Hz, higher ones are not observed and probably damped out.

\begin{figure}[ht] % matlab program:  plot_multichord_rib_bands_equal_dist_pvdf.m
\subfigure{(\textbf{a})}{\includegraphics[width=0.47\textwidth]{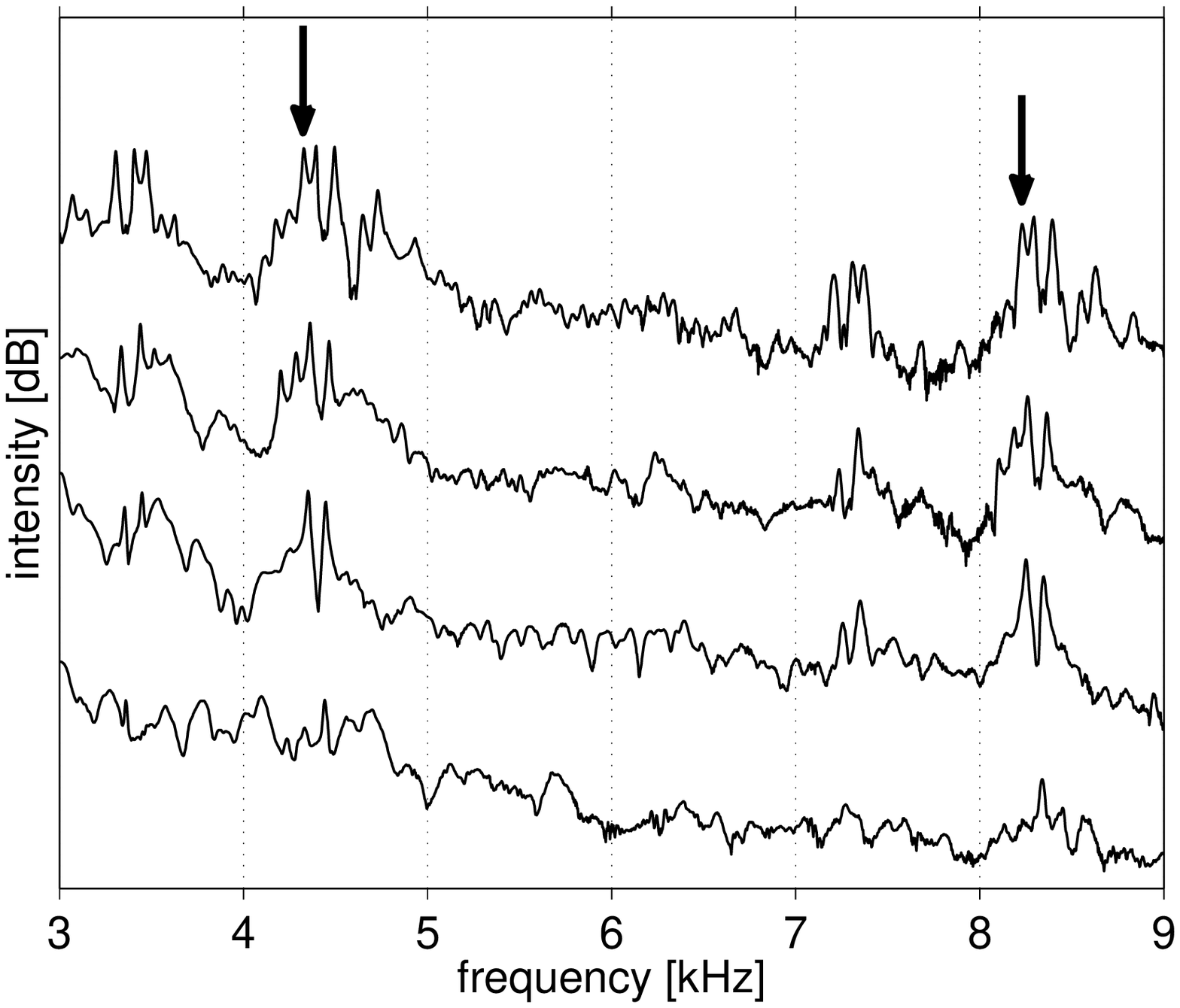}}  %% equal rib distance // increasing rib number // only for the PVDF // 14cm 
\subfigure{(\textbf{b})}{\includegraphics[width=0.47\textwidth]{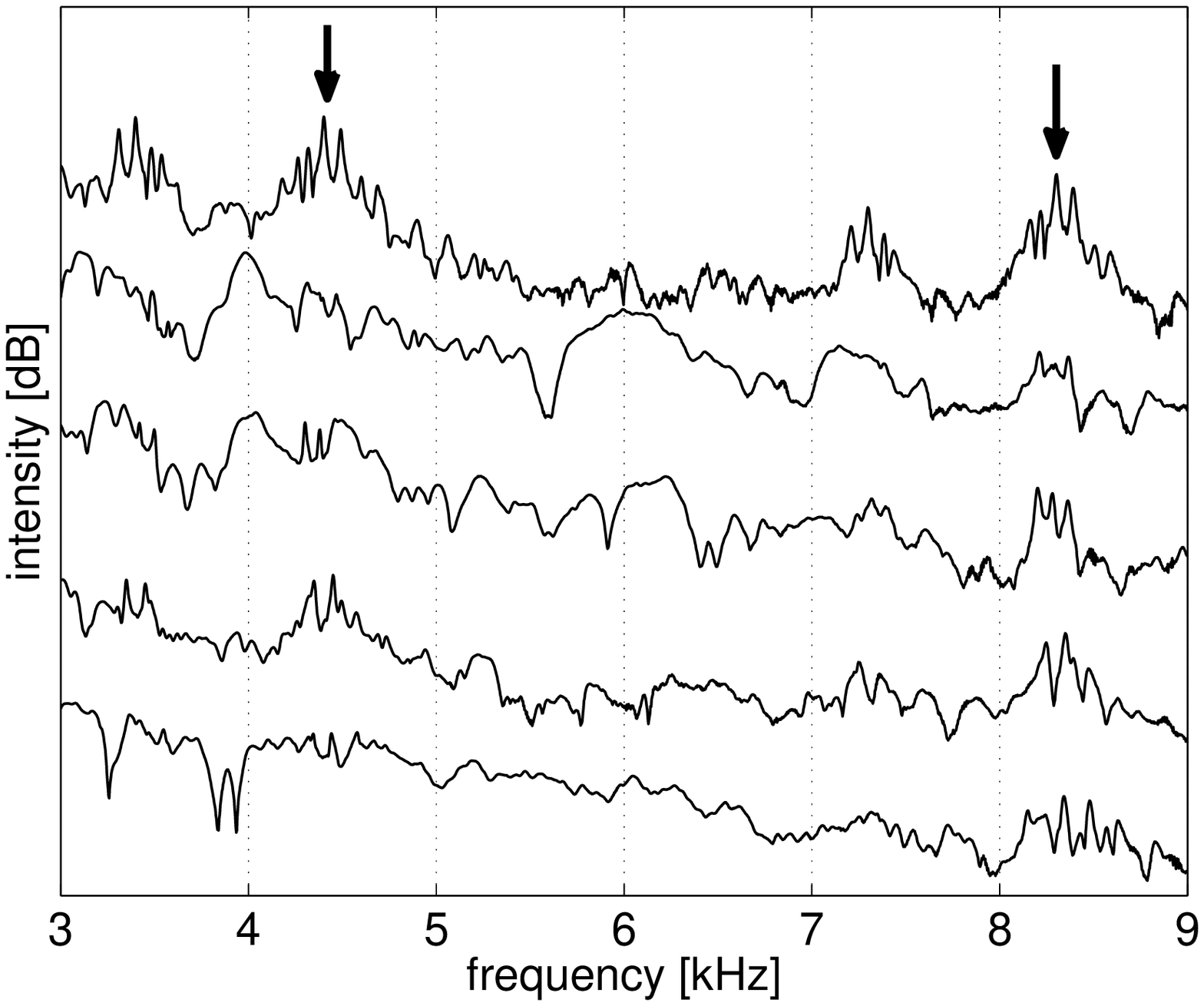}}  %% equal rib distance // increasing rib number // only for the PVDF // 10cm
\caption[Equal rib distance passbands.]{Inreasing number of ribs for different rib distances -- \textbf{a} with 14 cm and \textbf{b} with 9.8 cm distance. The number of ribs increases from zero at the bottom graph to the top, each graph by 2 (0-2-4-6-8 ribs). Only the first two pass bands are shown. i.e. the frequencies from  3000 Hz to 9000 Hz. The arrows indicate the positions of the developing pass bands. One can nicely recognize the evolution from 0 to 2,4,6 peaks in the band.
All the curves are displayed with a factor of 100 for better comparability. }
\label{fig:increasing_nr_of_ribs_equal_dist}
\end{figure}

Let us now discuss the effect of increasing number of ribs.
In Fig.~\ref{fig:increasing_nr_of_ribs_equal_dist} we show measurements for
a distance of 14 cm distance with increasing number of ribs.
Each new rib was placed 14 cm from the last one inserted, starting from 
the bridge. The maximum number of ribs was 6. From the above, we expect to see for each eigenmode of inter-rib distance a band-filter-like structure rising with a new peak for each rib we add. 

Indeed, we can count 2,4,6 peaks, marked by arrows in Fig.~\ref{fig:increasing_nr_of_ribs_equal_dist} for the second band. 
Clearly, this holds for all bands in the spectrum
This is a very nice demonstration of
the validity of our qualitative understanding of the system.
For other distances, the observation is confirmed, just with more peaks
for more ribs. What about the exact positions of the passing frequencies?
We find in the second band for 2, 4 and 6 ribs the frequencies $4254$ Hz and $4351$ Hz (2 ribs); $4202$ Hz, $4286$ Hz and $4361$ Hz (4 ribs); $4120$ Hz, $4175$ Hz, $4248$ Hz, $4327$ Hz and $4395$ Hz (6 ribs). The other expected peaks were not allocable from the data. 
From the naive theory,
we expect the positions $\nu_2+\nu_1+2\cdot\epsilon \sin^2 (k_j j/N) $.
A comparison shows only a qualitative coincidence whose deviation from theory can 
be attributed in the sketchy character of the theory. A quantitative 
comparison must include the details described above.

\begin{figure}[ht] % matlab program:  plot_multichord_rib_bands_equal_number_pvdf.m
\includegraphics[width=0.47\textwidth]{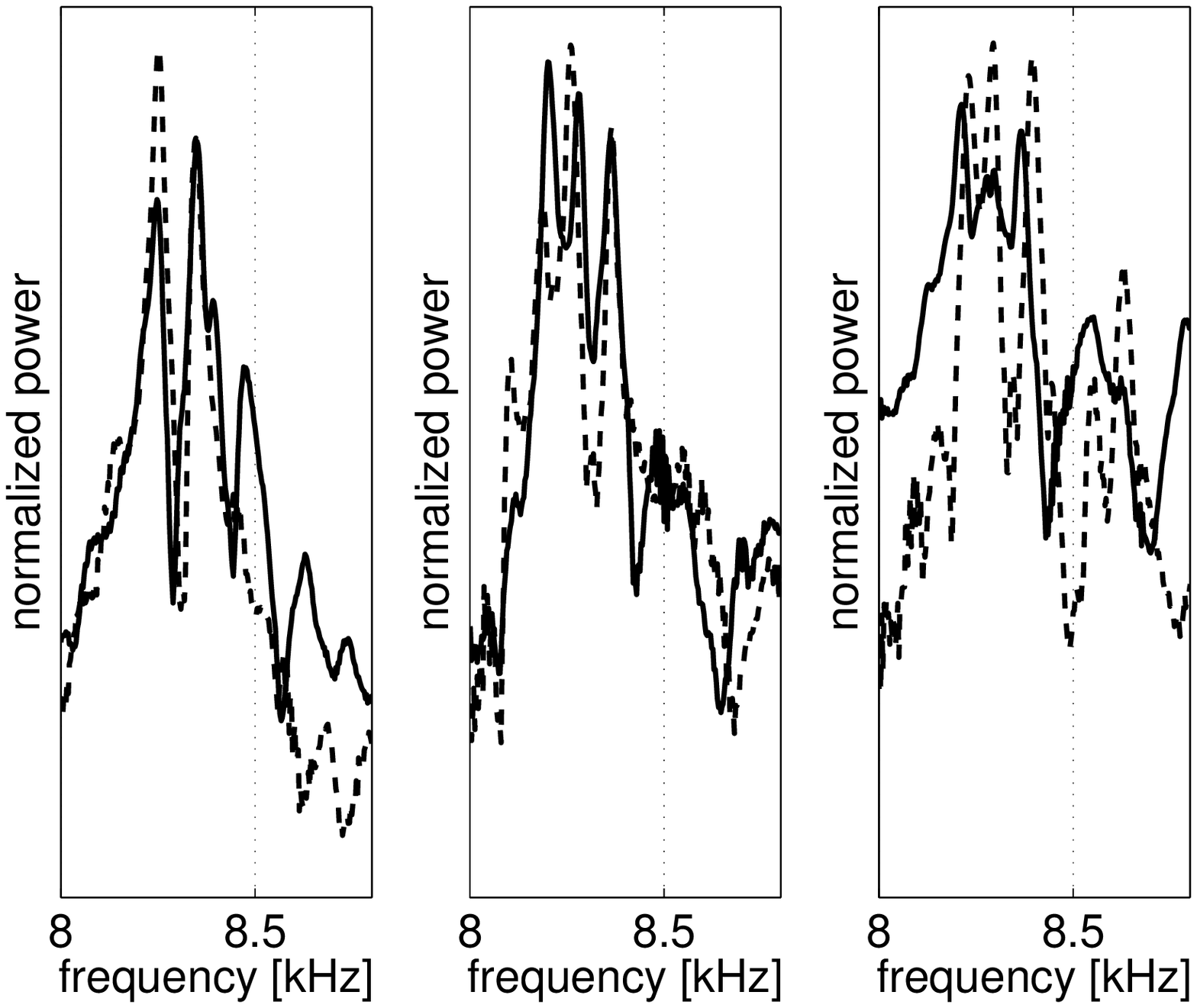}  %% equal rib number // only for the PVDF // 2 ribs
%\subfigure{(\textbf{b})}{\includegraphics[width=0.42\textwidth]{8b.eps}}  %% equal rib number // only for the PVDF // 4 ribs
%\subfigure{(\textbf{c})}{\includegraphics[width=0.42\textwidth]{8c.eps}}  %% equal rib number // only for the PVDF // 6 ribs
\caption[Equal rib number passbands.]{Comparison of the spectra with different rib distances but with equal rib numbers (2 ribs -- left; 4 ribs -- middle; 6 ribs -- right) for the piezopolymer measurement. The black dashed line represents the 14 cm, whereas the black solid line stands for the 9.8 cm rib distance. For each frame the third pass band in the frequency range of 8000 Hz to 8800 Hz is shown. Essentially only slight differences in the position of the peaks are found. Curves are normalized by total spectral power. }
\label{fig:equal_nr_different_distance}
\end{figure}

To measure what happens if we change the distance between ribs, we compare two 
measurements with an equal number of ribs, but two different distances. In Fig.~\ref{fig:equal_nr_different_distance}, we compare the measurements
for $d=14$ dm with $d=9.8$ cm distance, and the maximum possible number for 14 cm, namely 6 ribs.
We observe only a slight shift of the peaks, the band width and the band edges remain 
basically constant. Obviously the resonant frequencies $\frac{c}{2\cdot 0.14}\simeq 2760$ Hz, and $\frac{c}{0.098} \simeq 3943$ Hz do not play a role for the measured sound radiation and the internal soundboard vibrations dominate the spectral properties.

\begin{figure}[ht] % plot_multichord_rib_bands_time_dependent.m 
\includegraphics[height=0.47\textwidth]{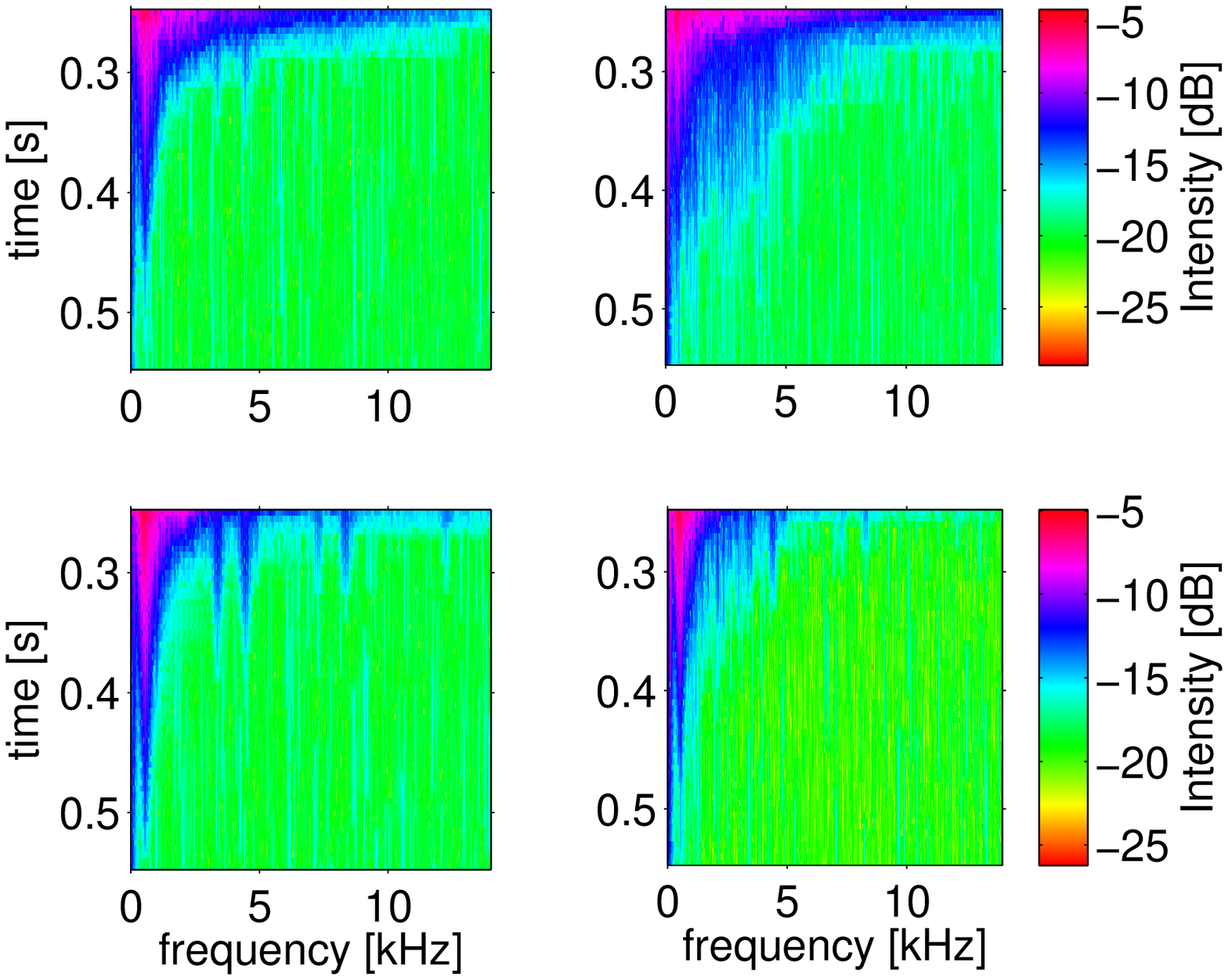} 
\caption[Time dependent frequency spectra]{Time dependent frequency spectrum of the resonance board with 0 ribs (\textbf{top}) and with the 13 ribs (\textbf{bottom}). The soundboard is excited once and the decay of the signal is observerd by the piezopolymer (\textbf{left}) and the microphone (\textbf{right}). One can see very clearly the development of the pass bands in the 13 ribs case (piezopolymer and microphone), where the characteristic is much more significant in the body vibration signal measured by the piezo film. For music radiated by such a board, a great change in the transmission properties results. Especially at  the microphone position the relative strength of the long-living frequencies has changed a lot. The band structure is not only visible by the measurements, but well hearable, as confirmed by the authors.}
\label{fig:nachhall}
\end{figure}

Finally, we want to discuss the effect on the reverberation/lifetime of modes. 
For musical purposes it is one of the most important issues, because 
this is what affects an audience listening to a piece of music played 
with an instrument of which the soundboard is a part.
To analyze it, we determined the spectrum with coarse resolution but for 
many short time slices of 5 ms each. Then the time record of these spectra
is plotted with a color coding for the spectral amplitudes. The result is 
shown in Fig.~\ref{fig:nachhall}.
For both, the field propagated in the soundboard (left part) and the radiated sound (right part)
a very inhomogeneous transfer function is found. For comparison, 
at the bottom the sound for the board without any ribs is shown, whereas the top displays the 
sound with ribs mounted. Again we use a configuration of 13 ribs. 
The difference suggests that the sound of the strings and 
hammer is heavily transformed by the mounted ribs. Not only that large parts of the 
spectrum are suppressed, as well the modes which are allowed to propagate
live much longer compared to the reference situation.

\section{Conclusion}
\label{sec:conclusions}

In this publication we demonstrate the dramatic effect obstacles mounted
on a soundboard may have on the radiated sound. We claim that the results hold
for any vibrating body with a regularly structured surface which radiates sound when
excited. With respect to impact  on musical performance, one has to evaluate the relevance for grand pianos sold and played nowadays, in particular with psychoacoustic analyses.
Of course, the geometry used is completely different and the modal structure is not as clear as on a rectangular soundboard, used here.
But from general arguments it follows that the spectrum for the grand piano has many similarities. On the other hand, musical instruments have
undergone a century-long optimization procedure and the complex interaction  of soundboard, bridge, string, and hammer certainly is optimized in a way that is most suitable to our hearing needs and the music played.

Nevertheless, we note that different manufacturers produce instruments
with different sound. Missing brilliance or,
vice versa, increased sound radiation for certain frequencies may be attributed 
to the geometry and restrictions in the construction of the soundboard which
acts as the last acoustical filter before sound is radiated towards 
an audience listening to a string instrument.

With respect to the measurement, we introduced a piezo-polymer measurement method to obtain information on the sound propagated from the bridge to the soundboard and within the soundboard itself. The advantage of this material is its extreme thin width which allows using it {\em inside} a setup without perturbing it. The radiated sound was measured with a conventional microphone. We obtained very good resolution with the piezopolymer, comparable with 
microphones. For the radiated sound, a systematic study over the full space could be helpful in verifying the results. We plan to use the piezopolymers 
in real instruments to transfer and verify our results for a grand piano. E.g. it is not clear in how far the different geometry of a grand piano soundboard 
changes the mode structure, e.g. it is possible that modes are pushed out of a band and enter the suppressed region, due to lateral reflections.

We used a specially designed setup with minimal parameter variations except from the mounting of the ribs to have full control over 
the reproducibility of the results. We varied the number of ribs mounted and the spacing between the ribs. as a result, we found 
that even few ribs heavily influence the sound radiated and propagated. The position of the passing frequencies, i.e. the eigenmodes
of the structured soundboard correspond with qualitative band theory, analogous to the theory for phonons in solid state physics. 

Now, the question arises how one can circumvent an unwanted suppression or increase of certain frequencies if nevertheless
ribs must be mounted on at soundboard to strengthen it mechanically. A structured band can be destroyed easily, if disorder
is introduced into the system. This is known from many investigations on Anderson localization \cite{Sheng-90,Bayer-Niederdraenk-93}.
The cost is a creation of localized modes which do not travel. The radiation from such inhomogeneous modes is a research topic on which 
to our knowledge no systematic results exist. It will be a very interesting topic of further research on radiation from
a soundboard with mounted structures. We conclude with noting that we touched a surprisingly rich research area, where musical instruments design might profit a lot from theory of solid-state physics.

% 
% In the first part we show that a prepared PVDF film can be used as a force calibrated sensor to record body vibrations in a music instrument. Important parameters were investigated with respect to the piezoelectric coefficient $d_{33}$ such as the initial pressure, the amplitude of the affecting force, the frequency response as well as the sensor long term stability. The parameter behavior provides the possibility of a calibrated body vibration measurement, which takes place on a simplified piano model (multichord). For the airborne sound a condenser microphone was placed above the arrangement. Additional ribs, like in a real piano, were placed on the rear side of the resonance board. Their number and distance variation results in spectral differences. A frequency band with its harmonics is observed in the spectrum, which can only be explained by the existence of these ribs. This effect becomes more prominent with an increasing rib number and a decreasing rib distance. 
% 
% 

\section*{Acknowledgements}
We thank R. Bader for fruitful discussion and encouragement.

\end{document}